\documentclass{article}
\usepackage{amsmath}
\usepackage{graphicx}
\usepackage{amsfonts}
\usepackage{amssymb}
\begin{document}

\begin{center}

\textbf{{\Large {The generalization of the addition property }}}

\smallskip

\textbf{{\Large {for soliton type processes}}}

\medskip

\textbf{Jerzy A. Zagrodzi\'{n}ski}

\medskip

\textit{ Institute of Physics, Polish Academy of Sciences,}

\textit{02-668 Warsaw, Poland; email: zagro@ifpan.edu.pl}
\end{center}

\bigskip

\textbf{Abstract}

A generalization of the addition relation for the Riemann theta functions and
its limiting version for exponential functions appearing in soliton type
equations are reported. The presented form seems to be particularly useful
when processes in $N+1,\;(N>1)$ space-time are analyzed. The commonly applied
bilinear and trilinear approaches, restricted to the pure soliton processes,
represent particular cases of the reported formalism. As an example, the
dispersion equation for either quasiperiodic or soliton processes following
2+1 Calogero-Bogoyavlenskij-Schiff equation is derived.

\medskip


\medskip

Running title: Generalization of addition property...

\bigskip

\section{INTRODUCTION}

As it is well known, solitons represent the localized excitations and appear
in the numerous branches of physics when nonlinear description is essential,
as in superconductivity, plasma physics, fiber optics, domain wall or fluid
dynamics, protein chains etc. Unfortunately, the well developed soliton theory
practically deals with the one-dimensional dynamical systems. The same
conclusion relates to the quasi-periodic processes. Because of the physical
reasons, many research groups make efforts in order to find the methods for
multidimensional solitons analysis. The results obtained hitherto have however
only a contributory character. Therefore below we present, it seems, a
slightly more general approach to the truly multidimensional solitons and
quasi periodic-processes.

In order to illustrate the problem we start with a simple example of the
famous Kadomtsev - Petviashvili equation%

\begin{equation}
3u_{yy}=\left[  4u_{t}-6uu_{x}+u_{xxx}\right]  _{x},\; \label{b1}%
\end{equation}
repeating some arguments of Dubrovin \cite{Du1}. Symbols $\left(  .\right)
_{x}\;$or $u_{x}$ denote relevant partial derivatives. Looking for the
solution in form%

\begin{equation}
\;u=u\left(  x,y,t\right)  =-2\ln\tau\left(  z_{1},...z_{g}\right)
,\;\;\;\;\;z_{i}=k_{i}x+l_{i}y+w_{i}t+z_{0,i},\;\;\;\;i=1,...,g\;, \label{b3}%
\end{equation}
a substitution to the equation (\ref{b1}) leads to
\begin{equation}
\frac{\tau_{xxxx}\tau-4\tau_{xxx}\tau_{x}+3\left(  \tau_{xx}\right)  ^{2}%
}{\tau^{2}}+4\frac{\tau_{xt}\tau-\tau_{x}\tau_{t}}{\tau^{2}}-3\frac{\tau
_{yy}\tau-\left(  \tau_{y}\right)  ^{2}}{\tau^{2}}=-8d, \label{b4}%
\end{equation}
if the first constant of integration is zero, and the second one is equal to
$-8d.$ According to the Hirota approach, there is introduced the bilinear
differential operator $D_{x}$ which if applied to the ordered pair of
functions $f\left(  x,y,t\right)  $ and $g\left(  x,y,t\right)  $ is defined as\ \cite{Hi}%

\begin{equation}
D_{x}^{N}\;\left(  f\circ g\right)  :=\left(  \partial_{w}\right)  ^{N}\left[
f\left(  x+w,y,t\right)  g\left(  x-w,y,t\right)  \right]  _{w=0}. \label{b5}%
\end{equation}
Then equation (\ref{b4}) can be written as
\begin{equation}
\left[  D_{x}^{4}+4D_{x}D_{t}-3\left(  D_{y}\right)  ^{2}\right]  \left(
\tau\circ\tau\right)  +8d\tau^{2}=0. \label{b6}%
\end{equation}
Next, for $d=0,$ making use the standard Hirota procedure one can find
multisoliton solutions of (\ref{b1}).

It is worth notice that the case $d\neq0$ is not considered in frame of the
bilinear technique since it leads to quasiperiodic solutions and periods of
the periodic subprocesses tend to infinity as $d\rightarrow0.$

There is an opinion that a huge success of the bilinear formalism in soliton
theory \cite{Hi}$^{,}$ \cite{Da} can be linked with the addition property for
$\tau$-functions.
\begin{equation}
\tau\left(  \mathbf{z}+\mathbf{w}\right)  \tau\left(  \mathbf{z}%
-\mathbf{w}\right)  =\sum_{\varepsilon}W\left(  \mathbf{w;\varepsilon}\right)
Z\left(  \mathbf{z;\varepsilon}\right)  , \label{p1}%
\end{equation}
where $\mathbf{z},\mathbf{w}\in\mathbb{C}^{g}$, $\tau:\mathbb{C}%
^{g}\rightarrow\mathbb{C},\;W,Z:\mathbb{C}^{g}\times\mathbb{Z}^{g}%
\rightarrow\mathbb{C}\;$, the sum is over a finite set and $Z\left(
\mathbf{z;\varepsilon}\right)  $ enumerated by $\mathbf{\varepsilon\in
}\mathbb{Z}_{2}^{g}$ form a set of linearly independent functions. Exponential
functions appearing in multisoliton solutions have this property as well as
the Riemann theta functions leading in turn to the quasiperiodic solutions.

The essential feature of the equation (\ref{p1}) is that the product of
shifted $\tau$-functions admits factorization.

Relation (\ref{p1}) should be considered only as the $\tau$ -function
property. This means that putting either $\mathbf{w}=0$ or $\mathbf{z}=0$ we
have $\tau^{2}\left(  \mathbf{z}\right)  =\sum_{\varepsilon}W\left(
0\mathbf{;\varepsilon}\right)  Z\left(  \mathbf{z;\varepsilon}\right)  $ or
$\tau\left(  \mathbf{w}\right)  \tau\left(  -\mathbf{w}\right)  =\sum
_{\varepsilon}W\left(  \mathbf{w;\varepsilon}\right)  Z\left(
\mathbf{0;\varepsilon}\right)  ,$ respectively. If the proper determinants do
not vanish, functions $W\left(  \mathbf{w;\varepsilon}\right)  $
\ and$\;Z\left(  \mathbf{z;\varepsilon}\right)  $ can be expressed by $\tau$-functions.

(Nota bene, if $\tau$-function is identified with the Riemann theta functions
$\theta\left(  z\right)  $ - see below, for $\mathbf{\varepsilon,\varepsilon
}^{\prime}\mathbf{\in}\mathbb{Z}_{2}^{g}$ , we have $\theta\left(
\mathbf{z}+\mathbf{\varepsilon}^{\prime}/2\right)  =\theta\left(
\mathbf{z}-\mathbf{\varepsilon}^{\prime}/2\right)  $, then $W\left(
\mathbf{\varepsilon}^{\prime}/2\mathbf{;\varepsilon}\right)  =\delta
_{\mathbf{\varepsilon,\varepsilon}^{\prime}}$ - Kronecker symbol and
\ $Z\left(  \mathbf{z;\varepsilon}\right)  =\theta^{2}\left(  \mathbf{z}%
+\mathbf{\varepsilon}/2\right)  .$)

Thus it seems that relation (\ref{p1}) is really very important and there are
a few practical reasons for such statement which we shall discuss below.

First of all, without any additional assumptions for function having addition
property one can easy calculate even derivatives of the equation (\ref{p1}),
obtaining some combinations of derivatives of $\tau\mathbf{-}$function
logarithm, \cite{JZ1}%

\begin{align}
2L_{ij}  &  =\sum_{\varepsilon}W_{w_{i}w_{j}}\left(  0\mathbf{;\varepsilon
}\right)  \;\left[  Z\left(  \mathbf{z;\varepsilon}\right)  /\tau^{2}\left(
\mathbf{z}\right)  \right]  ,\nonumber\\
& \label{p2}\\
L_{ijkl}+2\left(  3\times L_{ij}L_{kl}\right)   &  =\sum_{\varepsilon}%
W_{w_{i}w_{j}w_{k}w_{l}}\left(  0\mathbf{;\varepsilon}\right)  \;\left[
Z\left(  \mathbf{z;\varepsilon}\right)  /\tau^{2}\left(  \mathbf{z}\right)
\right]  ,\nonumber\\
& \label{p3}\\
L_{ijklmn}+2\left(  15\times L_{ij}L_{klmn}\right)  +4\left(  15\times
L_{ij}L_{kl}L_{mn}\right)   &  =\sum_{\varepsilon}W_{w_{i}w_{j}w_{k}w_{l}%
w_{m}w_{n}}\left(  0\mathbf{;\varepsilon}\right)  \;\left[  Z\left(
\mathbf{z;\varepsilon}\right)  /\tau^{2}\left(  \mathbf{z}\right)  \right]
,\nonumber\\
&  \label{p4}%
\end{align}
etc., where $L:=\ln\tau\left(  \mathbf{z}\right)  $, $L_{i}:=\partial_{z_{i}%
}\ln\tau\left(  \mathbf{z}\right)  \;$and we use a shorthand notation $\left(
3\times L_{ij}L_{kl}\right)  :=L_{ij}L_{kl}+L_{ik}L_{jl}+L_{il}L_{kj}$ i.e.
including all permutations.

Note that the ''basis'' functions $\left[  Z\left(  \mathbf{z;\varepsilon
}\right)  /\tau^{2}\left(  \mathbf{z}\right)  \right]  \;$for all operators
are the same.\ Therefore, the differentiation rules (\ref{p2}),\ (\ref{p3}),
(\ref{p4}), etc., automatically reconstruct the Korteweg - de Vries - Kotera -
Sawada hierarchy giving a tool for a derivation of relevant dispersion
equations \cite{Du1}$^{,}$ \cite{JZ1}.

In order to elucidate problem of dispersion equations, let us observe that
(\ref{b4}) can be written as%

\begin{equation}
L_{xxxx}+6\left(  L_{xx}\right)  ^{2}+4L_{xt}-3L_{yy}=-8d, \label{b11}%
\end{equation}
and, due to the linear dependence of arguments of $L=\ln\tau\left(
z_{1},...z_{g}\right)  $ on space and time coordinates, as%

\begin{equation}
\sum_{p,q,r,s=1}^{g}k_{p}k_{q}k_{r}k_{s}\left[  L_{z_{p}z_{q}z_{r}z_{s}%
}+2\left(  3\times L_{z_{p}z_{q}}L_{z_{r}z_{s}}\right)  \right]  +\sum
_{p,q=1}^{g}\left(  4k_{p}w_{q}-3l_{p}l_{q}\right)  L_{z_{p}z_{q}}=-8d.
\label{b12}%
\end{equation}
The shorthand notation $3\times...$ was explained before. Now, if $\tau-$
function has the addition property, we can apply rules (\ref{p2}) and
(\ref{p3}) obtaining
\begin{equation}
\sum_{\varepsilon}\left[
\begin{array}
[c]{c}%
\sum_{p,q,r,s=1}^{g}k_{p}k_{q}k_{r}k_{s}W_{w_{p}w_{q}w_{r}w_{s}}\left(
0\mathbf{;\varepsilon}\right)  +\\
+\sum_{p,q=1}^{g}\left(  4k_{p}w_{q}-3l_{p}l_{q}\right)  W_{w_{p}w_{q}}\left(
0\mathbf{;\varepsilon}\right)  +8dW\left(  0\mathbf{;\varepsilon}\right)
\end{array}
\right]  \;Z\left(  \mathbf{z;\varepsilon}\right)  =0. \label{b13}%
\end{equation}
since $\tau^{2}\left(  \mathbf{z}\right)  =\sum_{\varepsilon}W\left(
0\mathbf{;\varepsilon}\right)  Z\left(  \mathbf{z;\varepsilon}\right)
.\;$Finally, the Kadomtsev - Petviashvili equation (\ref{b1}) has a solution
if in a relevant class of $\tau-$functions (exponential or Riemann theta
function), for any $\mathbf{\varepsilon\in}\mathbb{Z}_{2}^{g}$ the system of
$2^{g}$ algebraic dispersion equations
\begin{equation}
\sum_{p,q,r,s=1}^{g}k_{p}k_{q}k_{r}k_{s}W_{w_{p}w_{q}w_{r}w_{s}}\left(
0\mathbf{;\varepsilon}\right)  +\sum_{p,q=1}^{g}\left(  4k_{p}w_{q}%
-3l_{p}l_{q}\right)  W_{w_{p}w_{q}}\left(  0\mathbf{;\varepsilon}\right)
+8dW\left(  0\mathbf{;\varepsilon}\right)  =0 \label{b14}%
\end{equation}
has a nontrivial solution. The broader discussion, particularly concerning
quasiperiodic solutions, one can find in \cite{Du1}$^{,}$ \cite{JZ1}$^{,}$
\cite{Du}$^{,}$ \cite{BE}\ . Here, we want to underline only a formal
similarity of (\ref{b6}) and (\ref{b14}) equations and to underline that
equation (\ref{b14}) remains valid also for$\;d\neq0,$ i.e. also for
quasiperiodic solutions

The second argument relates to another class of equations which are solved by
means of so-called trilinear operator. The trilinear operator $T$ and the
complex conjugate trilinear operator $T^{\ast}$ are defined as \cite{GRH}%
$^{,}$ \cite{YTF}
\begin{align}
T\left(  f\circ g\circ h\right)   &  =\left(  \partial_{z_{1}}+j\partial
_{z_{2}}+j^{2}\partial_{z_{3}}\right)  \;f\left(  z_{1}\right)  g\left(
z_{2}\right)  h\left(  z_{3}\right)  |_{z_{1}=z_{2}=z_{3}=z},\label{b15}\\
T^{\ast}\left(  f\circ g\circ h\right)   &  =\left(  \partial_{z_{1}}%
+j^{2}\partial_{z_{2}}+j\partial_{z_{3}}\right)  \;f\left(  z_{1}\right)
g\left(  z_{2}\right)  h\left(  z_{3}\right)  |_{z_{1}=z_{2}=z_{3}=z},
\label{b16}%
\end{align}
where $j=\exp\left(  i2\pi/3\right)  .$

Its application we illustrate by example of the Satsuma equation \cite{Sa}%

\begin{equation}
F_{xx}F_{yy}F-F_{xx}\left(  F_{y}\right)  ^{2}-F_{yy}\left(  F_{x}\right)
^{2}+2F_{xy}F_{x}F_{y}-\left(  F_{xy}\right)  ^{2}F=0, \label{b17}%
\end{equation}
which then can be written then as
\begin{equation}
\left(  T_{x}T_{x}^{\ast}T_{y}T_{y}^{\ast}-T_{x}^{\;2}T_{y}^{\ast2}\right)
\left(  F\circ F\circ F\right)  =0. \label{b18}%
\end{equation}

Observe now that the addition property for $F-$functions in version (\ref{p1})
is not applicable, since (\ref{b17}) consists triads of $F$ derivatives in
contrast to (\ref{b4}) where we have dealt with the pairs only. Therefore the
left hand side of a relevant version of addition property (\ref{p1}), if
exists, ought to contain the product of three shifted functions, (or even more).

The next argument relates to discrete equations and their reduction to
dispersion equations. For example, considering the KdV- type completely
integrable difference-difference equation%

\begin{align}
&  u\left(  x,t+d\right)  -u\left(  x,t-d\right)
\begin{tabular}
[c]{l}%
=
\end{tabular}
u\left(  x,t+d\right)  \;u\left(  x,t-d\right)  \;\left[  u\left(
x+d,t\right)  -u\left(  x-d,t\right)  \right]
,\;\;\;\;\;\;\;\;\;\;\;\label{p5}\\
&  u\left(  x,t\right)
\begin{tabular}
[c]{l}%
:=
\end{tabular}
\tau\left(  x+d,t\right)  \tau\left(  x-d,t\right)  /\tau\left(  x,t+d\right)
\tau\left(  x,t-d\right)  -1, \label{p6}%
\end{align}
it is seen that equation (\ref{p1}) enables to write the products $\tau\left(
x+d,t\right)  \tau\left(  x-d,t\right)  $ and

\noindent$\tau\left(  x,t+d\right)  \tau\left(  x,t-d\right)  $ in a compact
form leading to the dispersion equation \cite{JZ1}. The shift $\mathbf{w}=d$
plays a role of a step in a difference equation. On the other hand, if the
multidimensional version of difference-difference equation is considered, it
requires an introduction of a few and independent steps with respect to each
coordinate. This in a natural way leads to the form of addition property,
where on the left had side the product of a few shifted functions appear.

Now, one can ask about the class of functions having the property (\ref{p1}).
The oldest and the most known functions having this property are the Riemann
theta functions \cite{Ig}$^{,}$ \cite{JZ1}$^{,}$ \cite{Du}.
\begin{equation}
\theta\left(  \mathbf{z}|B\right)  =\sum_{n\in\mathbb{Z}^{g}}\exp\left[
i\pi\left(  2\left\langle \mathbf{z},\mathbf{n}\right\rangle +\left\langle
\mathbf{n},B\mathbf{n}\right\rangle \right)  \right]  \label{p7}%
\end{equation}
where $\mathbf{z}\in\mathbb{C}^{g},\;B\in\mathbb{C}^{g\times g}\;$is the
Riemann matrix, (i.e. symmetric with positively defined imaginary part),
$\left\langle \mathbf{z},\mathbf{n}\right\rangle :=\sum_{j=1}^{g}z_{j}n_{j}.$
The equivalent of (\ref{p1}) takes then the form
\begin{align}
&  \theta\left(  \mathbf{z}+\mathbf{w}|B\right)  \theta\left(  \mathbf{z}%
-\mathbf{w}|B\right)
\begin{tabular}
[c]{l}%
=
\end{tabular}
\label{p8}\\
&  =\sum_{\varepsilon\in\mathbb{Z}_{2}^{g}}\left[  \exp\left[  i\pi\left(
2\left\langle \mathbf{z,\varepsilon}\right\rangle +\left\langle
\mathbf{\varepsilon},B\mathbf{\varepsilon}\right\rangle \right)  \right]
\theta\left(  \mathbf{z}+B\mathbf{\varepsilon}|2B\right)  \exp\left(
i2\pi<2\mathbf{w},\mathbf{\varepsilon>}\right)  \theta\left(  \mathbf{w}%
+B\mathbf{\varepsilon}|2B\right)  \right] \nonumber
\end{align}
where symbol $\mathbf{\varepsilon}\in\mathbb{Z}_{2}^{g}$ $\ $denote \ g-fold
sum over $\varepsilon_{j}=0,1;$ $i=1,...,g.$ This relation can be also written
down by so called $\theta$ - functions with characteristics.

The second class is represented by exponential functions%

\begin{equation}
E\left(  \mathbf{z}|\tilde{B}\right)  =\sum_{n\in\mathbb{Z}_{2}^{g}}%
\exp\left[  i\pi\left(  2\left\langle \mathbf{z},\mathbf{n}\right\rangle
+\left\langle \mathbf{n},\tilde{B}\mathbf{n}\right\rangle \right)  \right]
\label{p9}%
\end{equation}
which appear in solutions of standard soliton equations. Matrix $\tilde{B}%
\in\mathbb{C}^{g\times g},$ (although sometimes it is convenient to assume
that diagonal elements of \ $\tilde{B}$ are real).\ In our opinion this is
just a source of the bilinear operator applicability. The combinations of
theta and exponential functions leading to processes when solitons propagate
on a quasiperiodic background form the next class of functions having addition
property \cite{JZ1}.

The fourth class, which to our knowledge has no physical application, is
generated by integral of product of Gaussian function and Riemann theta
functions with characteristics, where integration takes place with respect to
the second characteristics.

Concluding our motivation, the interpretation of the shift $\mathbf{w}$ in
(\ref{p1}) as a step suggests that for solitons in $N+1$ space-time a few
independent steps $\mathbf{w}_{1},$ ..., $\mathbf{w}_{N}$ should be
introduced. This leads to the generalization of the relation (\ref{p8}) or
more precisely of equation (\ref{p1}) and such generalization is just the aim
of this note. We believe that the reported below relations can be useful for
some multidimensional soliton type problems.

\section{THE\ GENERALIZED\ ADDITION\ PROPERTY}

In order to repeat the procedure as for (\ref{p2}-\ref{p4}) in case of a few
independent variables it would be sufficient to have relation%

\begin{equation}
\tau\left(  \mathbf{z}+\mathbf{u}^{\left(  0\right)  }\right)  \tau\left(
\mathbf{z}+\mathbf{u}^{\left(  1\right)  }\right)  ...\tau\left(
\mathbf{z}+\mathbf{u}^{\left(  J-1\right)  }\right)  =\sum_{\varepsilon
}W\left(  \mathbf{w}^{\left(  1\right)  },...,\mathbf{w}^{\left(  J-1\right)
};\mathbf{\varepsilon}\right)  Z\left(  \mathbf{z;\varepsilon}\right)
\label{p10}%
\end{equation}
for some class of $\tau$ -functions, where $\mathbf{u}^{\left(  j\right)
}=\mathbf{u}^{\left(  j\right)  }\left(  \mathbf{w}^{\left(  1\right)
},...,\mathbf{w}^{\left(  J\right)  }\right)  \in\mathbb{C}^{g},\;j=0,...,J-1$.

The reason is following: the derivatives of (\ref{p10}) with respect to
$w_{j}^{\left(  j\right)  }$ will relate only to $W-$function, leaving
$Z\left(  \mathbf{z;\varepsilon}\right)  $ unchanged. On the other hand the
derivatives of l.h.s.of (\ref{p10}) with respect to $w_{j}^{\left(  j\right)
} $ one can change into derivatives of $\tau-$ functions with respect $z_{i} $
which is necessary for an application in the soliton theory since usually
$\mathbf{z\;}$is linear in space and time variables, ($\mathbf{z:=k}%
_{x}x+\mathbf{k}_{y}y+...+\mathbf{k}_{z}z+\omega t\in\mathbb{C}^{g}).$ Since
as a $\tau$ -function, either exponential or Riemann theta functions are
usually chosen and since exponential functions can be considered as a
particular case of the second ones, we shall start from the Riemann theta functions.

There are numerous transformations for $\theta$ - functions, \cite{Ig}$^{,}$
\cite{Du}$^{,}$ \cite{BE}, but to our knowledge the $J$-th order addition
relation, as here, was considered only by Koizumi \cite{Ko} and cited in
\cite{BE}. But his factorization of the r.h.s. was completely different than
required in (\ref{p10}) and thus it is rather useless for our purposes.
Nevertheless for the standard $\theta$ - functions according to (\ref{p7}) one
can prove the identity which coincides with our demand (\ref{p10}), \cite{JAZ}.

{\large Theorem 1.}

If $\;\mathbf{z,u}^{\left(  k\right)  }\in\mathbb{C}^{g}${\Large ,}%
$\;k=0,..,J-1, $ such that $%
{\displaystyle\sum_{k=0}^{J-1}}
\mathbf{u}^{\left(  k\right)  }=0$, then%

\begin{align}
&  \theta\left(  \mathbf{z}+\mathbf{u}^{\left(  0\right)  }|B\right)
\;\theta\left(  \mathbf{z}+\mathbf{u}^{\left(  1\right)  }|B\right)
\;\theta\left(  \mathbf{z}+\mathbf{u}^{\left(  2\right)  }|B\right)
...\theta\left(  \mathbf{z}+\mathbf{u}^{\left(  J-1\right)  }|B\right)
\begin{tabular}
[c]{l}%
=
\end{tabular}
\nonumber\\
&  =\sum_{\varepsilon\in\mathbb{Z}_{J}^{g}}\exp\left[  i\pi\left(
2\left\langle \mathbf{z,\varepsilon}\right\rangle +\left\langle
\mathbf{\varepsilon},B\mathbf{\varepsilon}\right\rangle \right)  \right]
\;\theta\left(  J\mathbf{z}+B\mathbf{\varepsilon}|JB\right)  \;\times
\label{t2}\\
&  \times\exp\left(  i2\pi<\mathbf{u}^{\left(  0\right)  },\mathbf{\varepsilon
>}\right)  \;\;\theta\left(
\begin{array}
[c]{c}%
\mathbf{u}^{\left(  0\right)  }-\mathbf{u}^{\left(  1\right)  }%
+B\mathbf{\varepsilon}\\
\mathbf{u}^{\left(  0\right)  }-\mathbf{u}^{\left(  2\right)  }%
+B\mathbf{\varepsilon}\\
...\\
\mathbf{u}^{\left(  0\right)  }-\mathbf{u}^{\left(  J-1\right)  }%
+B\mathbf{\varepsilon}%
\end{array}
\left|  \left[
\begin{array}
[c]{llll}%
2B & B & .. & B\\
B & 2B & .. & B\\
.. & .. & 2B & ..\\
B & B & .. & 2B
\end{array}
\right]  \right.  \right) \nonumber
\end{align}
where all $\theta-$ functions are of order $g,$ with exception of the last one
on the r.h.s. which is of order $\left(  J-1\right)  g.$

The proof we report in Appendix 1.

Observe that the g-fold sum now is over $\varepsilon_{j}=0,1,...,J-1,$ (it
contains $J^{g}$ elements). If instead of $J$ vectorial parameters
$\mathbf{u}^{\left(  k\right)  },\;$among which only $J-1$ are independent,
the new $J-1$ vectorial parameters $\mathbf{w}^{\left(  k\right)  },\;\left(
k=1,...,J-1\right)  $ are introduced by%

\begin{equation}
\mathbf{u}^{\left(  k\right)  }=\left\{
\begin{array}
[c]{lll}%
\sum_{m=J-k+1}^{J-1}j^{m}\mathbf{w}^{\left(  m+k-J\right)  }+\sum_{m=1}%
^{J-k}j^{m}\mathbf{w}^{\left(  m+k-1\right)  }\;, & \text{for} &
k=1,...,J-1,\\
&  & \\
\sum_{k=1}^{J-1}\mathbf{w}^{\left(  k\right)  }\;, & \text{for} & k=0,
\end{array}
\right.  \label{t4}%
\end{equation}
where $j=\exp\left(  i2\pi/J\right)  ,\;$the relation $%
{\displaystyle\sum_{k=0}^{J-1}}
\mathbf{u}^{\left(  k\right)  }=0$ is satisfied automatically since
$\sum_{k=1}^{J-1}\mathbf{u}^{\left(  k\right)  }=-\sum_{k=1}^{J-1}%
\mathbf{w}^{\left(  k\right)  }.$ Thus equation (\ref{t2})\ with parameters
$\mathbf{u}_{k}$ given by (\ref{t4}) is satisfied for arbitrary set
$\mathbf{w}^{\left(  k\right)  }\in\mathbb{C}^{g}.$ The quantity $j$ is of
course $J$-th root from unity.

For fixed $J,$ equations (\ref{t4}) can be inverted, leading to%

\begin{equation}
\mathbf{w}^{\left(  k\right)  }=\frac{1}{j-1}\left(  \mathbf{u}^{\left(
k\right)  }-j\mathbf{u}^{\left(  k+1\right)  }\left(  1-\delta_{k,J-1}\right)
-j\mathbf{u}^{\left(  1\right)  }\delta_{k,J-1}\right)  ,\;\;\;\;k=1,...,J-1,
\label{t41}%
\end{equation}
where $\delta_{k,J-1}$ is the standard Kronecker symbol. Note that the choice
of $\mathbf{w}$ parameters as follows from (\ref{t4}) or (\ref{t41}) is not
unique. This one adopted here however, gives a correspondence with trilinear
operators introduced earlier in soliton theory, \cite{GRH}.

Applying a procedure denoted as the soliton limit \cite{JZ1} to the identity
(\ref{t2}) we are able present rewrite relations (\ref{t2}) for exponential
functions (\ref{p9}).

{\large Theorem 2.}%

\begin{align}
&  E\left(  \mathbf{z}+\mathbf{u}^{\left(  0\right)  }|\tilde{B}\right)
\;E\left(  \mathbf{z}+\mathbf{u}^{\left(  1\right)  }|\tilde{B}\right)
\;E\left(  \mathbf{z}+\mathbf{u}^{\left(  2\right)  }|\tilde{B}\right)
...E\left(  \mathbf{z}+\mathbf{u}^{\left(  J-1\right)  }|\tilde{B}\right)
\begin{tabular}
[c]{l}%
=
\end{tabular}
\label{t11}\\
&
\begin{tabular}
[c]{l}%
=
\end{tabular}
\sum_{\varepsilon\in\mathbb{Z}_{J}^{g}}\left\{  \exp\left[  i\pi2\left\langle
\left(  \mathbf{z+u^{\left(  0\right)  }}\right)  \mathbf{,\varepsilon
}\right\rangle \mathbf{+}\left\langle \mathbf{\varepsilon},\tilde
{B}\mathbf{\varepsilon}\right\rangle \right]  \times%
\begin{array}
[c]{c}%
\begin{array}
[c]{c}
\end{array}
\end{array}
\right. \nonumber\\
&  \left[  \sum_{m\in\mathbb{Z}_{2}^{g}}c\left(  \mathbf{\varepsilon
,m}\right)  \exp\left[  i\pi\left(  2\left\langle J\mathbf{z}%
+\mathbf{\varepsilon}\tilde{B},\mathbf{m}\right\rangle +J\left\langle
\mathbf{m},\tilde{B}\mathbf{m}\right\rangle \right)  \right]  \right]
\times\nonumber\\
&  \times\sum_{\mathbf{n}^{(1)},..,\mathbf{n}^{(J-1)}\in\mathbb{Z}_{2}^{g}%
}\left.  \left\{  C\left(  J,\mathbf{\varepsilon,n}\right)  \exp\left[
i2\pi\left(  \sum_{k=1}^{J-1}\left\langle -\left(  \mathbf{s}^{\left(
k\right)  }+\tilde{B}\mathbf{\varepsilon}\right)  ,\mathbf{n}^{(k)}%
\right\rangle +\sum_{k=1}^{J-1}\sum_{l=k}^{J-1}\left\langle \mathbf{n}%
^{(l)},\tilde{B}\mathbf{n}^{(k)}\right\rangle \right)  \right]  \;\right\}
\right\} \nonumber
\end{align}
where ''cut-off'' functions $c\left(  \mathbf{\varepsilon,m}\right)  \;$and
$C\left(  J,\mathbf{\varepsilon,n}\right)  \;$are given by
\begin{align}
c\left(  \mathbf{\varepsilon,m}\right)   &  =\prod_{j=0}^{g}\left(
\delta_{m_{j},0}+\delta_{m_{j},1}\delta_{\varepsilon_{j},0}\right)
;\;\;\;\;\;\label{t14}\\
C\left(  J,\mathbf{\varepsilon,n}\right)   &  =\prod_{j=1}^{g}\left[
\sum_{k=0}^{J-1}\delta_{\varepsilon_{j},k\;}\left(  \delta_{N_{j}\left(
J\right)  ,k-1}\left(  1-\delta_{k,0}\right)  +\delta_{N_{j}\left(  J\right)
,k}\right)  \right] \label{t15}\\
N_{j}\left(  J\right)   &  =\sum_{k=1}^{J-1}n_{j}^{\left(  k\right)
};\;\;\;\;\;\;\;\;\;\;\;\;\mathbf{s}^{\left(  k\right)  }=\mathbf{u}^{\left(
0\right)  }-\mathbf{u}^{\left(  k\right)  },\;\;\;k=1,..,J-1\;, \label{t16}%
\end{align}
$\mathbf{u}^{\left(  k\right)  }$ are defined by (\ref{t4}) and $\tilde{B}$
matrix is such that Diag $\operatorname{Im}$ $\tilde{B}=0,$ which however does
not infringe a generality.

The proof follows from the observation that exponential functions $E\left(
\mathbf{z}|\tilde{B}\right)  $ can be obtained from theta functions
$\theta\left(  \mathbf{z}|B\right)  $ as a limiting relation%

\[
E\left(  \mathbf{z}|\tilde{B}\right)  =\underset{all\;D_{ii}\rightarrow\infty
}{\lim}\theta\left(  \mathbf{z-}\frac{1}{2}iD\varepsilon_{0}|B\right)
,\;\;\;D:=Diag\operatorname{Im}\left(  B\right)  ,\;\;\;\tilde{B}%
:=B-iD,\;\;\varepsilon_{0}:=\left(  1,1,...,1\right)  ^{t},
\]
which was called - the soliton limit, \cite{JZ1}.

Similarly as in the case of standard addition property (\ref{p1}), one can
calculate some combinations of derivatives of $W$-function. They can be
reduced always to the different combinations of derivatives of $\tau$
-function logarithms, similarly as in case of the standard addition relation
(\ref{p2}-\ref{p4}). The sketch of procedure we present in Appendix 2.

\section{PARTICULAR\ CASES}

For $J=2,$ since $j=-1,\;$the identity (\ref{t2}) reduces to the commonly
known (\ref{p8}) formula, and similarly for exponential functions. For $J=3,$
since then $j=\exp\left(  i2\pi/3\right)  ,$ the identity (\ref{t2}) reduces
to
\begin{align}
&  \theta\left(  \mathbf{z+w}^{\left(  1\right)  }\mathbf{+w}^{\left(
2\right)  }|B\right)  \;\theta\left(  \mathbf{z}+j\mathbf{w}^{\left(
1\right)  }+j^{2}\mathbf{w}^{\left(  2\right)  }|B\right)  \;\theta\left(
\mathbf{z}+j^{2}\mathbf{w}^{\left(  1\right)  }+j\mathbf{w}^{\left(  2\right)
}|B\right)
\begin{tabular}
[c]{l}%
=
\end{tabular}
\label{jp3}\\
&
\begin{tabular}
[c]{l}%
=
\end{tabular}
\sum_{\varepsilon\in\mathbb{Z}_{3}^{g}}\;W\left(  \mathbf{w}^{\left(
1\right)  },\mathbf{w}^{\left(  2\right)  }\mathbf{;\varepsilon}\right)
Z\left(  \mathbf{z;\varepsilon}\right)  ,\nonumber\\
& \nonumber\\
&  W\left(  \mathbf{w}^{\left(  1\right)  },\mathbf{w}^{\left(  2\right)
}\mathbf{;\varepsilon}\right)
\begin{tabular}
[c]{l}%
=
\end{tabular}
\label{jp31}\\
&  \;\;
\begin{tabular}
[c]{l}%
=
\end{tabular}
\exp\left(  i2\pi<\left(  \mathbf{w}^{\left(  1\right)  }\mathbf{+w}^{\left(
2\right)  }\right)  ,\mathbf{\varepsilon>}\right)  \theta\left(
\begin{array}
[c]{c}%
\left(  1-j\right)  \mathbf{w}^{\left(  1\right)  }+\left(  1-j^{2}\right)
\mathbf{w}^{\left(  2\right)  }+B\mathbf{\varepsilon}\\
\left(  1-j^{2}\right)  \mathbf{w}^{\left(  1\right)  }+\left(  1-j\right)
\mathbf{w}^{\left(  2\right)  }+B\mathbf{\varepsilon}%
\end{array}
\left|  \left[
\begin{array}
[c]{ll}%
2B & B\\
B & 2B
\end{array}
\right]  \right.  \right) \nonumber\\
& \nonumber\\
&  Z\left(  \mathbf{z;\varepsilon}\right)
\begin{tabular}
[c]{l}%
=
\end{tabular}
\sum_{\varepsilon\in\mathbb{Z}_{3}^{g}}\exp\left[  i\pi\left(  2\left\langle
\mathbf{z,\varepsilon}\right\rangle +\left\langle \mathbf{\varepsilon
},B\mathbf{\varepsilon}\right\rangle \right)  \right]  \;\theta\left(
3\mathbf{z}+B\mathbf{\varepsilon}|3B\right)  . \label{jp32}%
\end{align}
where $\mathbf{z,\;w}^{\left(  1\right)  },\;\mathbf{w}^{\left(  2\right)
}\in C^{g}$ and are arbitrary. This relations rewritten for exponential
functions shows a close relation with trilinear operator analyzed in papers of
\cite{GRH}.

For $J=4,j=i$ and we obtain the product of four $\theta$ - functions in
version, which differs however from that reported by \cite{Ig} or \cite{BE}.

\section{AN EXAMPLE\ OF\ APPLICATION}

Let us consider the Calogero-Bogoyavlenskij-Schiff (CBS) equation $\left(
\eta=0\right)  ,$ which follows also from a reduction of self-dual Yang-Mills
equation and its modification by Grammaticos-Ramani-Hietarinta $\left(
\eta=1\right)  ,$ \cite{GRH}, of which soliton solutions were reported in
\cite{YTF}%
\begin{equation}
\Phi_{xt}+\frac{1}{4}\Phi_{xxxy}+\Phi_{x}\Phi_{xy}+\frac{1}{2}\Phi_{xx}%
\Phi_{y}+\eta\frac{1}{4}\partial_{x}^{-1}\Phi_{yyy}=0.\label{d1}%
\end{equation}
After substitution\ $\Phi=2L_{x}\;,\;$where$\;L=\ln\tau$ \ this equation
reduces to
\begin{equation}
\eta L_{yyy}+4L_{xxt}+L_{xxxxy}+8L_{xx}L_{xxy}+4L_{xy}L_{xxx}=C.\label{d3}%
\end{equation}
We assume that the integration constant $C$ is independent of $y$ and $t$ as
well. Moreover, if $\tau=\tau\left(  \mathbf{z}\right)  =\tau\left(
z_{1},...,z_{q}\right)  ,\;z_{k}=\kappa_{k}x+\nu_{k}y+\omega_{k}t,\;k=1,...,g$
, the equation (\ref{d3}) becomes $(L_{ijk}:=\partial_{z_{i}}\partial_{z_{j}%
}\partial_{z_{k}}\ln\tau\left(  \mathbf{z}\right)  \;)$%

\begin{align}
&  \sum_{ijk}\left(  \eta\nu_{i}\nu_{j}\nu_{k}+4\kappa_{i}\kappa_{j}\omega
_{k}\right)  L_{ijk}+\label{d6}\\
&  \;\;\;\;+\frac{1}{12}\sum_{ijklm}\left[  \left(  8\kappa_{i}\kappa
_{j}\kappa_{k}\kappa_{l}\nu_{m}+4\kappa_{i}\nu_{j}\kappa_{k}\kappa_{l}%
\kappa_{m}\right)  \left(  L_{ijklm}+12L_{ij}L_{klm}\right)  \right]
\begin{tabular}
[c]{l}%
=
\end{tabular}
C\;\nonumber
\end{align}

As the $\tau$-function we can choose either $\theta$ -functions or exponential
$E-$functions, (or even some combinations of both in spirit of those reported
in \cite{JZ1}).

On the other hand\ for $J=3,$ substituting $\mathbf{w}=\mathbf{w}^{\left(
1\right)  },\mathbf{v=w}^{\left(  2\right)  }$ in (\ref{jp3}) for typographic
reasons and denoting%

\begin{equation}
S\left(  \mathbf{z},\mathbf{w},\mathbf{v}\right)  =\tau\left(  \mathbf{z}%
+\mathbf{w}+\mathbf{v}\right)  \;\tau\left(  \mathbf{z}+j\mathbf{w}%
+j^{2}\mathbf{v}\right)  \;\tau\left(  \mathbf{z}+j^{2}\mathbf{w}%
+j\mathbf{v}\right)  , \label{d10}%
\end{equation}
simple but a little tedious calculations give the necessary relations%

\begin{align}
&  \partial_{w_{i}w_{j}w_{k}}S\left(  \mathbf{z},\mathbf{w},\mathbf{v}\right)
|_{\mathbf{w}=\mathbf{v=0}}/S\left(  \mathbf{0},\mathbf{0},\mathbf{0}\right)
=\partial_{v_{i}v_{j}v_{k}}S\left(  \mathbf{z},\mathbf{w},\mathbf{v}\right)
|_{\mathbf{w}=\mathbf{v=0}}/S\left(  \mathbf{0},\mathbf{0},\mathbf{0}\right)
=3L_{ijk}\label{d12}\\
&  \partial_{w_{i}w_{j}w_{k}w_{l}v_{m}}S\left(  \mathbf{z},\mathbf{w}%
,\mathbf{v}\right)  |_{\mathbf{w}=\mathbf{v=0}}/S\left(  \mathbf{0}%
,\mathbf{0},\mathbf{0}\right)  =3\left[  L_{ijklm}+12L_{ij}L_{klm}\right]  .
\label{d13}%
\end{align}

Derivatives of (\ref{jp3}) with respect to $w$ and $v$ deal with the $W$
function and since $S\left(  \mathbf{z},\mathbf{0},\mathbf{0}\right)  =\left[
\theta\left(  z|B\right)  \right]  ^{3},$ we have (see Appendix 2)
\begin{align}
L_{ijk}  &  =\frac{1}{3}\sum_{\varepsilon\in\mathbb{Z}_{3}^{g}}\;\;\;\left[
W_{w_{i}w_{j}w_{k}}\left(  \mathbf{w},\mathbf{v;\varepsilon}\right)
|_{\mathbf{w}=\mathbf{v=0}}\;Z\left(  \mathbf{z;\varepsilon}\right)  \right]
/\theta^{3}\left(  z|B\right) \label{d18}\\
L_{ijklm}+12L_{ij}L_{klm}  &  =\frac{1}{3}\left[  \sum_{\varepsilon
\in\mathbb{Z}_{3}^{g}}W_{w_{i}w_{j}w_{k}w_{l}v_{m}}\left(  \mathbf{w}%
,\mathbf{v;\varepsilon}\right)  |_{\mathbf{w}=\mathbf{v=0}}\;Z\left(
\mathbf{z;\varepsilon}\right)  \right]  /\theta^{3}\left(  z|B\right)  .
\label{d19}%
\end{align}

Coefficients $W\left(  \mathbf{w},\mathbf{v;\varepsilon}\right)  $ and
functions $Z\left(  \mathbf{z;\varepsilon}\right)  $ are given now either by
(\ref{jp31}) or (\ref{jp32}), respectively, if quasiperiodic solutionsare
considered or by their soliton limit in spirit of (\ref{t11}).

Considering functions $Z\left(  \mathbf{z;\varepsilon}\right)  $ as
independent, (\ref{d6}) reduces to the system of $3^{g}$ algebraic dispersion
equations for any $\varepsilon\in\mathbb{Z}_{3}^{g}$%

\begin{align}
&  \sum_{ijk}\left(  \eta\nu_{i}\nu_{j}\nu_{k}+4\kappa_{i}\kappa_{j}\omega
_{k}\right)  W_{w_{i}w_{j}w_{k}}\left(  \mathbf{w},\mathbf{v;\varepsilon
}\right)  |_{\mathbf{w}=\mathbf{v=0}}+\label{d20}\\
&  \sum_{ijklm}\kappa_{i}\kappa_{j}\kappa_{k}\kappa_{l}\nu_{m}W_{w_{i}%
w_{j}w_{k}w_{l}v_{m}}\left(  \mathbf{w},\mathbf{v;\varepsilon}\right)
|_{\mathbf{w}=\mathbf{v=0}}=3CW\left(  \mathbf{w},\mathbf{v;\varepsilon
}\right)  |_{\mathbf{w}=\mathbf{v=0}}\;\nonumber
\end{align}
which determines relations between $\kappa_{i},\nu_{i},\omega_{i}$ $(i=1...g)$
and also $C,$ if $\eta=1$. Its nontrivial solution, if exists, gives the
quasiperiodic solutions of 2+1 CBS equations in form of $\Phi=2\left[
\ln\theta\left(  \mathbf{z}|B\right)  \right]  _{x}$ with $z_{i}=\kappa
_{i}x+\nu_{i}y+\omega_{i}t$. For higher genus $g,$ it gives also the
additional conditions on matrix $B$, since system (\ref{d20}) can be
overdetermined. then).

Multisoliton solutions of this equation, reported in paper \cite{YTF}, were
found by means of the trilinear operator formalism, where the third roots
$\sqrt[3]{1}\;$appear as above, but they can be also derived using equation
(\ref{t11}) taken for $J=3$. There is an interesting similarity between
trilinear soliton version of CBS equation reported in \cite{YTF} and the left
hand side of equation (\ref{d20}) which when $C\neq0$ is valid also for
quasi-periodic processes. It follows from the fact that trilinear operators
$T$ and $T^{\ast}$ used in papers \cite{GRH} and \cite{YTF} coincide with used
here operators $\partial/\partial w$ and $\partial/\partial v,$ respectively.

In conclusion, it seems that the presented here generalized addition formula
can be useful for other $N+1$ soliton type equations.

\section{ACKNOWLEDGMENTS}

The author acknowledges helpful discussions with M. Jaworski and a support of
The Polish Committee for Scientific Researches (KBN), Grants No. 2P03B-114-11
and 2P03B-148-14.\ Special thanks belong to T. Nikiciuk for numerous relations
calculated as in Appendix 2.

\newpage

\section{APPENDIX 1}

\textbf{The proof of Theorem1}.

Lemat 1

If $B$ is Riemannian then also matrix $B^{\prime}=\left[
\begin{array}
[c]{lll}%
2B & .. & B\\
.. & 2B & ..\\
B & .. & 2B
\end{array}
\right]  $ ( with diagonal blocks $2B$ and otherwise $B)$ is also Riemannian.
Indeed, both matrices $B\;$and\ $B^{\prime}$are symmetric. Moreover, if
$\left\langle \mathbf{n,Bn}\right\rangle \geq0,$ for any $\mathbf{n}\in Z^{g}
$ (with equality for $\mathbf{n=0),}$ then
\[
\left\langle \left(  \mathbf{n}_{1}\mathbf{,...,n}_{J-1}\right)
\mathbf{,B}^{\prime}\left(  \mathbf{n}_{1}\mathbf{,...,n}_{J-1}\right)
^{T}\right\rangle =\sum_{k=1}^{J-1}\left\langle \mathbf{n}_{k}\mathbf{,Bn}%
_{k}\right\rangle +\left\langle \left(  \sum_{k=1}^{J-1}\mathbf{n}_{k}\right)
\mathbf{,B}\left(  \sum_{k=1}^{J-1}\mathbf{n}_{k}\right)  \right\rangle
\geq0,
\]
(with equality for all $\mathbf{n}_{k}\mathbf{=0}$).

By the definition of $\theta$-function (\ref{p7}), the l.h.s of (\ref{t2})
takes form%

\begin{align}
&  \sum_{\mathbf{n}^{\left(  0\right)  }\in\mathbb{Z}^{g}}...\sum
_{\mathbf{n}^{\left(  J-1\right)  }\in\mathbb{Z}^{g}}\exp\left[  i\pi\left(
2\sum_{j=0}^{J-1}\left\langle \left(  \mathbf{z+u}^{\left(  j\right)
}\right)  ,\mathbf{n}^{\left(  j\right)  }\right\rangle +\sum_{j=0}%
^{J-1}\left\langle \mathbf{n}^{\left(  j\right)  }B\mathbf{n}^{\left(
j\right)  }\right\rangle \right)  \right]  =\label{1a1}\\
&  =\sum_{\mathbf{n}^{\left(  0\right)  }\in\mathbb{Z}^{g}}...\sum
_{\mathbf{n}^{\left(  J-1\right)  }\in\mathbb{Z}^{g}}\exp\left[  i\pi\left(
\begin{array}
[c]{c}%
\left[  2\left\langle \left(  \mathbf{z+u}^{\left(  0\right)  }\right)
,\mathbf{n}^{\left(  0\right)  }\right\rangle +\left\langle \mathbf{n}%
^{\left(  0\right)  },B\mathbf{n}^{\left(  0\right)  }\right\rangle \right]
+\\
+2\sum_{j=1}^{J-1}\left\langle \left(  \mathbf{z+u}^{\left(  j\right)
}\right)  ,\mathbf{n}^{\left(  j\right)  }\right\rangle +\sum_{j=1}%
^{J-1}\left\langle \mathbf{n}^{\left(  j\right)  },B\mathbf{n}^{\left(
j\right)  }\right\rangle
\end{array}
\right)  \right]  \label{1a2}%
\end{align}

Substituting $\mathbf{n}=\mathbf{n}^{\left(  0\right)  }-\sum_{j=1}%
^{J-1}\mathbf{n}^{\left(  j\right)  }$,we have%

\begin{align}
&  =\sum_{\mathbf{n}\in\mathbb{Z}^{g}}\;\sum_{\mathbf{n}^{\left(  1\right)
},,\mathbf{n}^{\left(  J-1\right)  }\in\mathbb{Z}^{g}}\exp\left[  i\pi\left(
\begin{array}
[c]{c}%
2\left[  \left\langle \left(  \mathbf{z+u}^{\left(  0\right)  }\right)
,\left(  \mathbf{n}-\sum_{i=1}^{J-1}\mathbf{n}^{\left(  i\right)  }\right)
\right\rangle +\sum_{j=1}^{J-1}\left\langle \left(  \mathbf{z+u}^{\left(
j\right)  }\right)  ,\mathbf{n}^{\left(  j\right)  }\right\rangle \right]  +\\
\left\langle \left(  \mathbf{n}-\sum_{i=1}^{J-1}\mathbf{n}^{\left(  i\right)
}\right)  ,B\left(  \mathbf{n}-\sum_{i=1}^{J-1}\mathbf{n}^{\left(  i\right)
}\right)  \right\rangle +\sum_{j=1}^{J-1}\left\langle \mathbf{n}^{\left(
j\right)  },B\mathbf{n}^{\left(  j\right)  }\right\rangle
\end{array}
\right)  \right] \nonumber\\
&  \label{1a50}%
\end{align}%

\begin{align}
&  =\sum_{\mathbf{n}\in\mathbb{Z}^{g}}\;\sum_{\mathbf{n}^{\left(  1\right)
},,\mathbf{n}^{\left(  J-1\right)  }\in\mathbb{Z}^{g}}\exp\left[  i\pi\left(
\begin{array}
[c]{c}%
2\left[  \left\langle \left(  \mathbf{z+u}^{\left(  0\right)  }\right)
,\mathbf{n}\right\rangle -\sum_{i=1}^{J-1}\left\langle \left(  \mathbf{u}%
^{\left(  0\right)  }-\mathbf{u}^{\left(  j\right)  }\right)  ,\mathbf{n}%
^{\left(  i\right)  }\right\rangle \right]  +\\
+\left\langle \left(  \mathbf{n}-\sum_{i=1}^{J-1}\mathbf{n}^{\left(  i\right)
}\right)  ,B\left(  \mathbf{n}-\sum_{i=1}^{J-1}\mathbf{n}^{\left(  i\right)
}\right)  \right\rangle +\sum_{j=1}^{J-1}\left\langle \mathbf{n}^{\left(
j\right)  },B\mathbf{n}^{\left(  j\right)  }\right\rangle
\end{array}
\right)  \right] \nonumber\\
&  \label{1a41}%
\end{align}%

\begin{equation}
=\sum_{\mathbf{n}\in\mathbb{Z}^{g}}\;\sum_{\mathbf{n}^{\left(  1\right)
},,\mathbf{n}^{\left(  J-1\right)  }\in\mathbb{Z}^{g}}\exp\left[  i\pi\left(
\begin{array}
[c]{c}%
2\left[  \left\langle \left(  \mathbf{z+u}^{\left(  0\right)  }\right)
,\mathbf{n}\right\rangle -\sum_{i=1}^{J-1}\left\langle \left(  \mathbf{u}%
^{\left(  0\right)  }-\mathbf{u}^{\left(  j\right)  }+B\mathbf{n}\right)
,\mathbf{n}^{\left(  i\right)  }\right\rangle \right]  +\\
+\left\langle \mathbf{n},B\mathbf{n}\right\rangle +\left\langle \sum
_{i=1}^{J-1}\mathbf{n}^{\left(  i\right)  },B\sum_{i=1}^{J-1}\mathbf{n}%
^{\left(  i\right)  }\right\rangle +\sum_{j=1}^{J-1}\left\langle
\mathbf{n}^{\left(  j\right)  },B\mathbf{n}^{\left(  j\right)  }\right\rangle
\end{array}
\right)  \right]  \label{1a51}%
\end{equation}%

\begin{align}
&  =\sum_{\mathbf{n}\in\mathbb{Z}^{g}}\exp i\pi\left[  2\left\langle \left(
\mathbf{z+u}^{\left(  0\right)  }\right)  ,\mathbf{n}\right\rangle
+\left\langle \mathbf{n},B\mathbf{n}\right\rangle \right]  \times
\label{1a60}\\
&  \;\;\;\;\;\;\;\;\;\;\;\;\;\;\times\sum_{\mathbf{n}^{\left(  1\right)
},.,\mathbf{n}^{\left(  J-1\right)  }\in\mathbb{Z}^{g}}\exp\left[
i\pi\left(
\begin{array}
[c]{c}%
2\left[  \sum_{i=1}^{J-1}\left\langle \left(  \mathbf{u}^{\left(  j\right)
}-\mathbf{u}^{\left(  0\right)  }-B\mathbf{n}\right)  ,\mathbf{n}^{\left(
i\right)  }\right\rangle \right]  +\\
+\left\langle \sum_{i=1}^{J-1}\mathbf{n}^{\left(  i\right)  },B\sum
_{i=1}^{J-1}\mathbf{n}^{\left(  i\right)  }\right\rangle +\sum_{j=1}%
^{J-1}\left\langle \mathbf{n}^{\left(  j\right)  },B\mathbf{n}^{\left(
j\right)  }\right\rangle
\end{array}
\right)  \right] \nonumber\\
& \nonumber\\
&  =\sum_{\mathbf{n}\in\mathbb{Z}^{g}}\exp i\pi\left[  2\left\langle \left(
\mathbf{z+u}^{\left(  0\right)  }\right)  ,\mathbf{n}\right\rangle
+\left\langle \mathbf{n},B\mathbf{n}\right\rangle \right]  \theta\left(
\left[
\begin{array}
[c]{c}%
\mathbf{u}^{\left(  1\right)  }-\mathbf{u}^{\left(  0\right)  }-B\mathbf{n}\\
\mathbf{...}\\
\mathbf{u}^{\left(  J-1\right)  }-\mathbf{u}^{\left(  0\right)  }-B\mathbf{n}%
\end{array}
\right]  \left|  \left[
\begin{array}
[c]{lll}%
2B & .. & B\\
.. & 2B & ..\\
B & .. & 2B
\end{array}
\right]  \right.  \right)  .\nonumber\\
&  \label{1a72}%
\end{align}

The next substitution is $\mathbf{n}=J\mathbf{m+\varepsilon,}$ with
$\mathbf{n}\in\mathbb{Z}^{g}$ , $\mathbf{\varepsilon}\in\mathbb{Z}_{J}^{g}$
and thus $\sum_{\mathbf{n}\in\mathbb{Z}^{g}}=\sum_{\mathbf{m}\in\mathbb{Z}%
^{g}}\sum_{\mathbf{\varepsilon}\in\mathbb{Z}_{J}^{g}}$ , due to standard
properties of theta functions \cite{Du1},\cite{BE} that%

\[
\theta\left(  \mathbf{z}+A\mathbf{m|}A\right)  =\exp\left\{  -i\pi\left[
2\left\langle \mathbf{z},\mathbf{m}\right\rangle +\left\langle \mathbf{m}%
,A\mathbf{m}\right\rangle \right]  \right\}  \theta\left(  \mathbf{z|}%
A\right)  ,\;\;\text{if\ \ \ }\mathbf{z\in}\mathbb{C}\mathbf{^{n}%
,\;\;}A\mathbf{\in}\mathbb{C}\mathbf{^{n\times n}\;}\text{and\ \ \ }%
\mathbf{m}\in\mathbb{Z}^{n},
\]
the expression (\ref{1a72}) becomes

\bigskip\
\begin{align}
&  =\sum_{\mathbf{m}\in\mathbb{Z}^{g}}\sum_{\mathbf{\varepsilon}\in
\mathbb{Z}_{J}^{g}}\exp\left\{  i\pi\left[  2\left\langle \left[
\mathbf{z+u}^{\left(  0\right)  }\right]  ,\left(  J\mathbf{m+\varepsilon
}\right)  \right\rangle +\left\langle \left(  J\mathbf{m+\varepsilon}\right)
,B\left(  J\mathbf{m+\varepsilon}\right)  \right\rangle \right]  \right\}
\times\label{1a11}\\
&  \;\;\;\;\;\;\;\;\;\times\theta\left(  \left[
\begin{array}
[c]{c}%
\mathbf{u}^{\left(  0\right)  }-\mathbf{u}^{\left(  1\right)  }+B\left(
J\mathbf{m+\varepsilon}\right) \\
\mathbf{...}\\
\mathbf{u}^{\left(  0\right)  }-\mathbf{u}^{\left(  J-1\right)  }+B\left(
J\mathbf{m+\varepsilon}\right)
\end{array}
\right]  \left|  \left[
\begin{array}
[c]{lll}%
2\mathbf{B} & .. & \mathbf{B}\\
.. & .. & ..\\
\mathbf{B} & .. & 2\mathbf{B}%
\end{array}
\right]  \right.  \right) \nonumber
\end{align}%
\begin{align}
&  =\sum_{\mathbf{m}\in\mathbb{Z}^{g}}\sum_{\mathbf{\varepsilon}\in
\mathbb{Z}_{J}^{g}}\exp\left\{  i\pi\left[  \left\langle 2\left(
\mathbf{z+u}^{\left(  0\right)  }\right)  +B\left(  J\mathbf{m+\varepsilon
}\right)  ,\left(  J\mathbf{m+\varepsilon}\right)  \right\rangle \right]
\right\}  \times\label{1a130}\\
&  \;\;\;\;\;\;\;\;\times\theta\left(  \left[
\begin{array}
[c]{c}%
\mathbf{u}^{\left(  0\right)  }-\mathbf{u}^{\left(  1\right)  }%
+B\mathbf{\varepsilon}\\
\mathbf{...}\\
\mathbf{u}^{\left(  0\right)  }-\mathbf{u}^{\left(  J-1\right)  }%
+B\mathbf{\varepsilon}%
\end{array}
\right]  +\left[
\begin{array}
[c]{lll}%
2\mathbf{B} & .. & \mathbf{B}\\
.. & .. & ..\\
\mathbf{B} & .. & 2\mathbf{B}%
\end{array}
\right]  \left[
\begin{array}
[c]{c}%
\mathbf{m}\\
\mathbf{..}\\
\mathbf{m}%
\end{array}
\right]  \left|  \left[
\begin{array}
[c]{lll}%
2\mathbf{B} & .. & \mathbf{B}\\
.. & .. & ..\\
\mathbf{B} & .. & 2\mathbf{B}%
\end{array}
\right]  \right.  \right) \nonumber
\end{align}%
\begin{align}
&  =\sum_{\mathbf{\varepsilon}\in\mathbb{Z}_{J}^{g}}\exp\left[  i\pi
\left\langle 2\left(  \mathbf{z+u}^{\left(  0\right)  }\right)  +B\left(
J\mathbf{m+\varepsilon}\right)  ,\left(  J\mathbf{m+\varepsilon}\right)
\right\rangle \right]  \times\nonumber\\
&  \;\;\;\;\;\;\;\;\;\times\exp\left\langle -i\pi\left[  2\left\langle
{\textstyle\sum_{j=1}^{J-1}}
\left(  \mathbf{u}^{\left(  0\right)  }-\mathbf{u}^{\left(  j\right)
}+B\mathbf{\varepsilon}\right)  \mathbf{,m}\right\rangle +\left\langle
\mathbf{m}J\left(  J-1\right)  B\mathbf{m}\right\rangle \right]  \right\rangle
\times\label{1a14}\\
&  \;\;\;\;\;\;\;\;\times\theta\left(  \left[
\begin{array}
[c]{c}%
\mathbf{u}^{\left(  0\right)  }-\mathbf{u}^{\left(  1\right)  }%
+B\mathbf{\varepsilon}\\
\mathbf{...}\\
\mathbf{u}^{\left(  0\right)  }-\mathbf{u}^{\left(  J-1\right)  }%
+B\mathbf{\varepsilon}%
\end{array}
\right]  \left|  \left[
\begin{array}
[c]{lll}%
2\mathbf{B} & .. & \mathbf{B}\\
.. & .. & ..\\
\mathbf{B} & .. & 2\mathbf{B}%
\end{array}
\right]  \right.  \right) \nonumber\\
& \nonumber\\
&  =\sum_{\mathbf{\varepsilon}\in\mathbb{Z}_{J}^{g}}\exp\left\{  i\pi\left[
2\left\langle \left(  \mathbf{z+u}_{0}\right)  ,\mathbf{\varepsilon
}\right\rangle +\left\langle \mathbf{\varepsilon},B\mathbf{\varepsilon
}\right\rangle \right]  \right\}  \theta\left(  J\mathbf{z+}%
{\textstyle\sum_{j=0}^{J-1}}
\mathbf{u}^{\left(  j\right)  }+B\mathbf{\varepsilon|}JB\right)
\times\label{1a16}\\
&  \;\;\;\;\;\;\;\;\times\;\theta\left(  \left[
\begin{array}
[c]{c}%
\mathbf{u}^{\left(  0\right)  }-\mathbf{u}^{\left(  1\right)  }%
+\mathbf{B\varepsilon}\\
\mathbf{...}\\
\mathbf{u}^{\left(  0\right)  }-\mathbf{u}^{\left(  J-1\right)  }%
+\mathbf{B\varepsilon}%
\end{array}
\right]  \left|  \left[
\begin{array}
[c]{lll}%
2\mathbf{B} & .. & \mathbf{B}\\
.. & .. & ..\\
\mathbf{B} & .. & 2\mathbf{B}%
\end{array}
\right]  \right.  \right) \nonumber
\end{align}
i.e. we have obtained the right hand-side of (\ref{t2}) under condition that $%
{\textstyle\sum_{j=0}^{J-1}}
\mathbf{u}^{\left(  j\right)  }=0$. \rule{5pt}{5pt}

\newpage

\section{APPENDIX 2}

\textbf{Algorithm for derivatives of W-function.}

The $W$-function derivatives and thus the combination of derivatives of $\tau
$-function leading to some hierarchy of pde in N+1 space-time can be derived
algorithmically using e.g. the Mathematica program.

Starting from the more useful form of (\ref{p10})%

\begin{equation}
\exp\sum_{j=1}^{J}\left[  \ln\tau\left(  \mathbf{z}+\mathbf{u}^{\left(
j\right)  }\right)  -\ln\tau\left(  \mathbf{z}\right)  \right]  =\sum
_{\varepsilon}W\left(  \mathbf{w}^{\left(  1\right)  },...,\mathbf{w}^{\left(
J-1\right)  };\mathbf{\varepsilon}\right)  \frac{Z\left(
\mathbf{z;\varepsilon}\right)  }{\left[  \tau\left(  \mathbf{z}\right)
\right]  ^{J}} \label{t20}%
\end{equation}
with $\mathbf{u}^{\left(  k\right)  }=\mathbf{u}^{\left(  k\right)  }\left(
\mathbf{w}^{\left(  1\right)  },...,\mathbf{w}^{\left(  J-1\right)  }\right)
\;$and differentiating with respect to the components of $\mathbf{w}^{\left(
k\right)  }$ vectors we obtain the expressions for combinations of derivatives
of $\tau$ -function logarithms. A simple algorithm for this calculation we
illustrate here giving two examples.%

\begin{align}
J  &  =3;\;\;j=\exp\left(  i2\pi/J\right)  =\exp\left(  i2\pi/3\right)
;\;\;\mathbf{w}^{\left(  k\right)  }\in C^{g},\;\;k=1,2\nonumber\\
&  \frac{\partial^{p+q}}{\left(  \partial w^{\left(  1\right)  }\right)
^{p}\left(  \partial w^{\left(  2\right)  }\right)  ^{q}}W\left(
\mathbf{w}^{\left(  1\right)  },\mathbf{w}^{\left(  2\right)  }\right)
\begin{tabular}
[c]{l}%
=
\end{tabular}
\label{t21}\\
&  =\frac{\partial^{p+q}}{\left(  \partial w^{\left(  1\right)  }\right)
^{p}\left(  \partial w^{\left(  2\right)  }\right)  ^{q}}\exp\left[  L\left(
\mathbf{w}^{\left(  1\right)  }+\mathbf{w}^{\left(  2\right)  }\right)
+L\left(  j\mathbf{w}^{\left(  1\right)  }+j^{2}\mathbf{w}^{\left(  2\right)
}\right)  +L\left(  j^{2}\mathbf{w}^{\left(  1\right)  }+j\mathbf{w}^{\left(
2\right)  }\right)  -3L\left(  \mathbf{0}\right)  \right]  ,\nonumber
\end{align}%

\begin{align}
J  &  =5;\;\;j=\exp\left(  i2\pi/J\right)  =\exp\left(  i2\pi/5\right)
;\;\;\mathbf{w}^{\left(  k\right)  }\in C^{g},\;\;k=1,...,4\nonumber\\
&  \frac{\partial^{p+q}}{\left(  \partial w^{\left(  1\right)  }\right)
^{p}\left(  \partial w^{\left(  2\right)  }\right)  ^{q}}W\left(
\mathbf{w}^{\left(  1\right)  },...,\mathbf{w}^{\left(  4\right)  }\right)
\begin{tabular}
[c]{l}%
=
\end{tabular}
\label{t22}\\
&  =\frac{\partial^{p+q}}{\left(  \partial w^{\left(  1\right)  }\right)
^{p}\left(  \partial w^{\left(  2\right)  }\right)  ^{q}}\exp\left[
\begin{array}
[c]{c}%
L\left(  \;\;\mathbf{w}^{\left(  1\right)  }\;+\;\;\mathbf{w}^{\left(
2\right)  }\;+\;\;\mathbf{w}^{\left(  3\right)  }\;+\;\text{\ }\mathbf{w}%
^{\left(  4\right)  }\right)  +\\
L\left(  j\mathbf{w}^{\left(  1\right)  }+j^{2}\mathbf{w}^{\left(  2\right)
}+j^{3}\mathbf{w}^{\left(  3\right)  }+j^{4}\mathbf{w}^{\left(  4\right)
}\right)  +\\
L\left(  j^{4}\mathbf{w}^{\left(  1\right)  }+j\mathbf{w}^{\left(  2\right)
}+j^{2}\mathbf{w}^{\left(  3\right)  }+j^{3}\mathbf{w}^{\left(  4\right)
}\right)  +\\
L\left(  j^{3}\mathbf{w}^{\left(  1\right)  }+j^{4}\mathbf{w}^{\left(
2\right)  }+j\mathbf{w}^{\left(  3\right)  }+j^{2}\mathbf{w}^{\left(
4\right)  }\right)  +\\
L\left(  j^{2}\mathbf{w}^{\left(  1\right)  }+j^{3}\mathbf{w}^{\left(
2\right)  }+j^{4}\mathbf{w}^{\left(  3\right)  }+j\mathbf{w}^{\left(
4\right)  }\right)  -\\
5L\left(  \mathbf{0}\right)
\end{array}
\right] \nonumber
\end{align}

where have used a shorthand notation $L\left(  \mathbf{w}\right)  :=\ln
\tau\left(  \mathbf{z+w}\right)  .$ \


\newpage

\begin{thebibliography}{99}
\bibitem{Du1}{\footnotesize B. A. Dubrowin, Dispersion Relations for Nonlinear
Waves and Schottky Problem, in: Important Developments in Soliton Theory, eds.
A.S.Fokas and V.E.Zakharov, (Springer, Berlin. 1993).}

\bibitem {Hi}{\footnotesize R. Hirota, Phys. Rev. Lett. 27, 1192 (1971); also
in B\"{a}cklund Transformations p.40, eds. A.Dodd and B. Eckmann, (Springer,
Berlin, 1976); also J. Phys. Soc. Japan, 43, 1424 (1977); in Solitons, eds. R.
K. Bullough, P. J. Caudrey, (Springer, Berlin, 1980); }

\bibitem {Da}{\footnotesize E. Date et al.,Transformation groups for soliton
eqns., in: Nonlinear integrable systems - classical and quantum theory, eds.
M. Jimbo and T. Miwa, (World Scientific, Singapore 1983).}

\bibitem {JZ1}{\footnotesize J. Zagrodzi\'{n}ski, J. Phys. A 15, 3109 (1982);
Phys. Rev. E, 51, 2566 (1995); also in: NEEDS'94, eds. V. G. Makhankov, A. R.
Bishop, D. Holm, (World Scientific, Singapore, 1995).}

\bibitem {Du}{\footnotesize B. A. Dubrowin, Usp. Mat. Nauk, 36, 12 (1980), (in Russian).}

\bibitem {BE}{\footnotesize E. D. Bielokolos, A. I. Bobenko, V. Z. Enolskii,
A. R. Its, V. B. Matveev, Algebro-Geometric Approach to Nonlinear Integrable
Equations, (Springer, Berlin, 1994).}

\bibitem {GRH}{\footnotesize B. Grammaticos, A. Ramani, J. Hietarinta, Phys.
Lett. A, 190, 65 (1994); also in NEEDS'94, ed. V. G. Makhankov, A. R. Bishop,
D. Holm, (World Scientific, Singapure, 1995).}

\bibitem {YTF}{\footnotesize S. J. Yu, K. Toda, T. Fukuyama, J. Phys. A 31,
3337 (1998); J. Phys. A 31, 10181 (1998). }

\bibitem {Sa}{\footnotesize J. Matsukidaira, J. Satsuma and W. Strampp,
Phys.Lett.A, 147,467 (1990).}

\bibitem {Ig}{\footnotesize J. Igusa, Theta Functions, (Springer, Berlin, 1972).}

\bibitem {Ko}{\footnotesize S. Koizumi, Math. Ann. 242, 127 (1979).}

\bibitem {JAZ}{\footnotesize J.A.Zagrodzi\'{n}ski, Repts on Math. Phys. 46,
311 (2000); also Proc. of Workshop '' Nonlinearity, Integrability and all of
that: 20-years after NEEDs'79'' eds. M.Boiti et all. p. 226, (World
Scientific, Singapure, 2000)}
\end{thebibliography}
\end{document}